# An efficient iterative method for looped pipe network hydraulics

Dejan Brkić [1,*] and Pavel Praks [2,*]

[1] Research and Development Center "Alfatec", 18000 Niš, Serbia
[2] IT4Innovations, VŠB - Technical University of Ostrava, 708 00 Ostrava, Czech Republic

* Correspondence: dejanrgf@tesla.rcub.bg.ac.rs or dejanbrkic0611@gmail.com, https://orcid.org/0000-0002-2502-0601 (D.B.); pavel.praks@vsb.cz or pavel.praks@gmail.com, https://orcid.org/0000-0002-3913-7800 (P.P.)

**Abstract:** Original and improved version of the Hardy Cross iterative method with related modifications are today widely used for calculation of fluid flow through conduits in loops-like distribution networks of pipes with known node fluid consumptions. Fluid in these networks is usually natural gas for distribution in the municipalities, water in waterworks or hot water in district heating system, air in the case of ventilation systems in buildings or mines, etc. Since, the resistances in these networks depend of flow, problem is not linear like in electrical circuits, and iterative procedure must be used. In both version of the Hardy Cross method, in original and in the improved one, initial results of calculation in iteration procedure is not flow, but rather the correction of flow. Unfortunately, these corrections should be added to or subtracted from a flow calculated in previous iteration according to complicate algebraic rules. After the here presented node-loop method, final results in each of the iterations is flow directly rather than flow correction. In that way complex algebraic scheme for sign of flow correction is avoided, while the final results still remain unchanged. Numbers of required iterations for the same results are comparable with the improved Hardy Cross method.

## 1. Introduction

Since, the resistances in a network of pipes for distribution of fluids depend on flow, problem is not linear like in electric circuits and iterative procedure must be used to calculate distribution of fluid flow through pipes and distribution of pressure in the network. Usually, in a hydraulic network of pipes, consumption of fluid assigned to each node is known and stays unchanged during computation. This is also the case for the inputs in network which are also assigned to nodes and which also do not change during calculation. Further, in order to calculate flow and pressure distribution in the network of pipes, first of all, initial flow pattern through pipes in the network has to be assigned to satisfy first Kirchhoff law for each node. This means to satisfy material balance of fluid moved through network. During iterative cycles of calculation, this flow distribution will changes in order to conform second prerequisite condition govern by the second Kirchhoff law, i.e. to satisfy energy balance in each closed conduit formed by pipes in the network. In hydraulic network this energy balance is usually expressed through pressure or some of the functions in which pressure exist. While the first Kirchhoff law has to be satisfied in all iterations for each node in the network, the second Kirchhoff law has to be satisfied for each closed conduit at the end of calculation.

Usually, such as in Hardy Cross method [1] and related improved version [2], result of iterative calculation of flow distribution pattern in a hydraulic network is correction of flow [1-3]. This correction of flow has to be added to flow calculated in the previous iteration using complex algebraic rules [3,4]. This intermediate step will be eliminated, using procedure that will be shown in this paper. In that way, flow will be directly calculated in all iteration for each pipe.

All methods from this paper assume equilibrium between pressure and friction forces in steady and incompressible flow. As a result, they cannot be successfully used in unsteady and compressible flow calculations with large pressure drop where inertia force is important. Gas flow in a municipal distribution network [5], air flow in a ventilation system in buildings and mines [6], and of course water flow in waterworks [7] or district heating systems [8] and cooling systems [8] can be treated as incompressible flow since the pressure drop in these kinds of networks are minor even to compress significantly natural gas or air. The same applies to pipelines for distribution of mixed natural gas and hydrogen [9].

## 2. Overview of existing methods for calculation of flow distribution in a looped network of pipes

*2.1. Loop-oriented methods; Original and improved Hardy Cross method*

The Hardy Cross method [1] introduced in 1936 is the first useful procedure for the calculation of flow distribution in looped networks of pipes. Further step was made by introduction of the modification in the original Hardy Cross method in 1970 by Epp and Fowler [2]. The original Hardy Cross method [1] as a sort of single adjustment method, first of all, as an intermediate step in calculation, determines correction of flow for each loop independently and then applies this corrections to compute new flow in each conduit. It is not efficient as the improved Hardy Cross method [2,3] that considers entire system simultaneously. The improved Hardy Cross method [2], still firstly as an intermediate step, determines corrections for each loop but treated all network system simultaneously, and then applies this correction to compute new flow in each conduit such as in the original version [1]. It is more efficient, but also intermediate step in calculation is not eliminated. More than thirty years had to pass by before the introduction of the modification by Epp and Fowler [2] only because of matrix calculation. While use of matrix form in the original Hardy Cross method is not mandatory [1], for the improved version it is [2]. In the original paper of Hardy Cross from 1936 [1], problem is not solved using any kind of matrix calculation (but also this approach can be expressed using matrix calculation with no affects on final results [7]).

*2.2. Node-oriented methods*

Two years before modification of the original Hardy Cross method, Shamir and Howard in 1968 [10] reformulated original method to solve node equations and not any more loop equations like in the original Hardy Cross method [1]. The node equations expressed in the node method in terms of unknown pressure in nodes [11]. Methods based on node equations are less reliable which means that the single adjustment methods based on idea from the original Hardy Cross method (but here adjust for nodes) must be employed with caution. Idea for these node-oriented methods is simple knowing principle of loop-oriented method developed by Hardy Cross [1]. In a loop-oriented method, energy distribution for all closed paths in a network governs by the second Kirchhoff law will be always satisfied, while material balance for all nodes in a network governs by the first Kirchhoff law will be balanced in an iterative procedure. Similar principle applies as in the original Hardy Cross method, but only with opposite approach. Still, as an intermediate step, correction of pressure has to be calculated [12-14] (in the original method by Hardy Cross this is correction of flow [15-17]), and then after that, pressure as a final result of iteration has to be calculated using complex algebraic rules. Pressure can be expressed in different quantities, such lengths of water elevation or similar.

*2.3. Node-loop oriented method*

After the development of the loop-oriented and node-oriented methods, and after introduction of matrix calculus, all necessary tools are available, i.e. matrix form of loop method and matrix form of node method to unite both, the loop and the node equations in matrix form which has a result completely new and innovated method [18,19]. This transformation makes possible direct calculation of final flow in each of the iterations, and not the correction of flow like in methods mentioned before (Figure 1). Unfortunately, as already explained these corrections of flow calculated after previous methods should be added to or subtracted from flow (or pressure in the node method) calculated in previous iterations according to complicated algebraic rules [3].

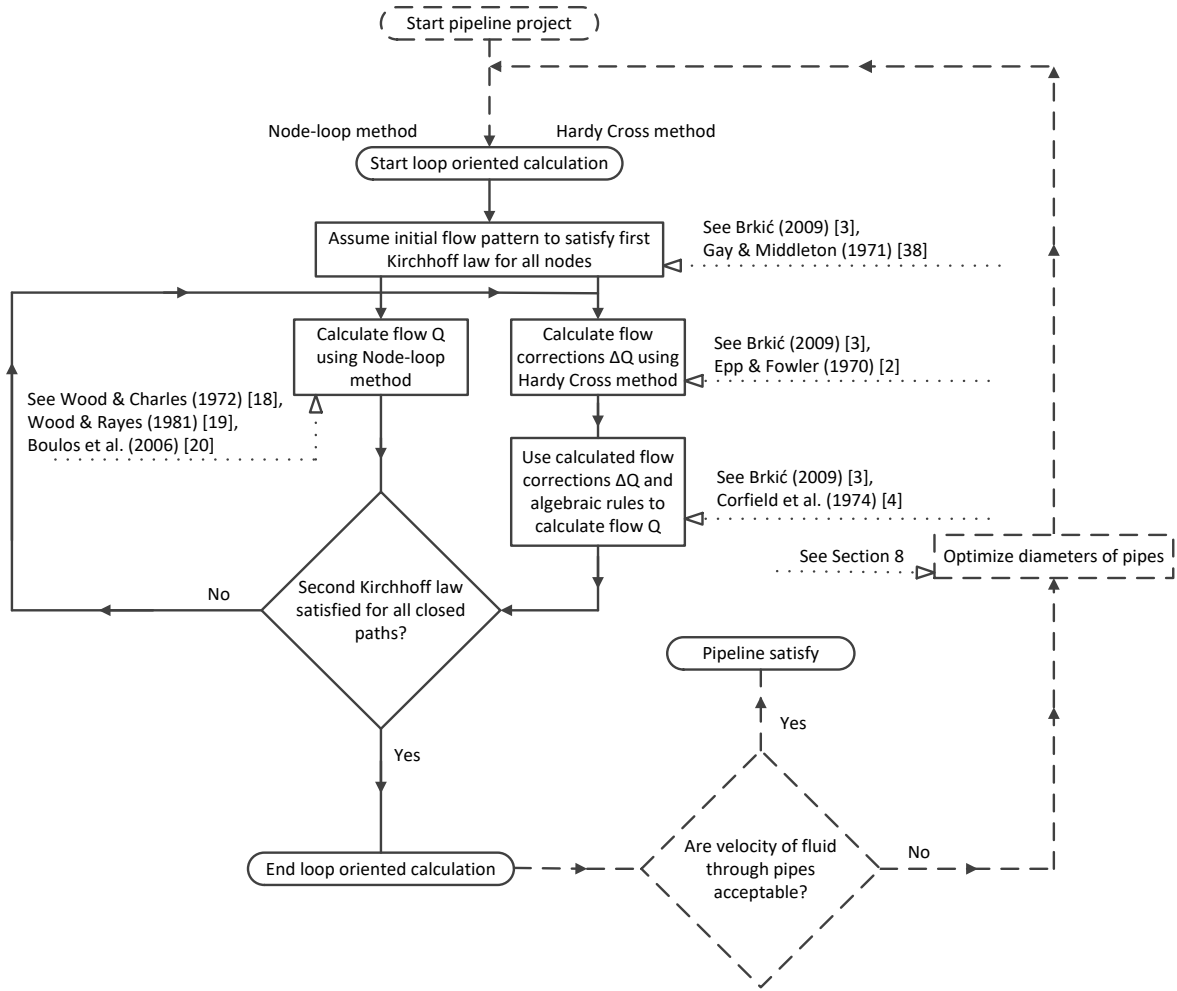

**Figure 1.** Main strength of the node-loop method compared with Hardy Cross is in direct flow calculation

So, the main strength of the node-loop method introduced in 1972 by Wood and Charles [18] for waterworks calculation does not reflect in noticeably reduced number of iteration compared to the modified Hardy Cross method. Main advantage of this method is in the capability to solve directly the pipe flow rate rather than flow correction. The method uses a linear head loss term which allows a network of n pipes to be described by a set of n linear equations which can be solved simultaneously for the flow distribution. Wood and Rayes in 1981 introduced improvement in the node-loop method [19]. Here will be shown improved version of this method rearranged for gas flow and for water flow in terms of pressure distribution rather than head distribution (which quantity is express in length; such as elevation of water).

## 3. Some literary overview of the existed methods for calculation of flow distribution in a looped network of pipes

Excellent example of calculation of looped natural gas distribution network after original Hardy Cross method can be found in Gas Engineers Handbook from 1974 [4]. Already mentioned algebraic rules for correction of flow calculated as an intermediate step in iterative procedure that can be used for both versions of Hardy Cross method can be found in this reference book [4] (and also for the node-oriented method but where correction of pressure is calculated as an intermediate step rather than correction of flow). This algebraic rules were further additionally and developed in Brkić [3]. Same spatial gas network as shown in Brkić [3] will be also used here for calculation after the node-loop method. Same topology of the network with same diameter will be used here for calculation of water flow as comparisons of the results obtained for liquid flow.

Excellent book in this issue, but only for waterworks calculation by Boulos et al. [20] can be recommended for further reading. In this book, unfortunately, the Hazen-Williams equation, an obsolete relation is used to correlate only water flow, pressure drops in pipes and hydraulics frictions.

Further, for details on natural ventilation airflow networks one can consult paper of Aynsley [6]. There is no space here to calculate separately air ventilation network, but readers interested in this matter can make this in a very effective way, according to natural gas and water flow calculation shown in this paper. Specific details on airflow resistances are also given in Aynsley [6].

Also, Todini and Pilati [21] for water networks and Hamam and Brameller [22] for gas networks wrote conservation of energy for each pipe and as result beside of flow correction in each pipe, pressure drop also can be simultaneously calculated. This method is also known as hybrid or gradient approach. Some comparisons of available methods for pipeline network calculations can be found in Mah [23], Mah and Shacham [24], Mah and Lin [25], etc. To compare calculation of water networks using the Hazen-Williams equation and approach with pseudo-loops consult book of Boulous et al [20]. Lopes [26] also deals with the program for the Hardy Cross solution of the piping networks. Shown kind of problems today can be solved very easily using MS Excel [27,28].

The first computer solutions of network problems were done on analog computers where electrical elements are used to simulate pipe networks [29]. Today, this approach is obsolete. Also, today natural gas is mostly distributed in cities, but earlier it was gas derived from coal [30].

## 4. Hydraulics resistance of a single pipe

Source-issue that cause problem with the calculation of hydraulic networks is non-constant value of hydraulic resistance when fluid convey through pipe. On the other hand, electrical resistance of a wire or a resistor has a constant value which has a consequence, non-iterative calculation of electrical circuits. To establish relation between flow rate of natural gas through a single pipe and related pressure drop, the Renouard equation for gas flow will be used and in that case (1) [25]. Using that approach, resistance will not be calculated at all since Renouard's equation relates pressure and flow rate using other properties, parameters and quantities to connect these two variables. On the other hand, for the calculation of hydraulic resistance in a single pipe, well known Colebrook equation will be used [26] (which is also iterative and which caused also some problems [33-35]) where pressure drop is calculated using Darcy-Weisbach equation. Finally, for calculation of air-flow through ventilation system, one can consult Aynsley [6], as already mentioned before.

The Hazen–Williams equation, which is used in here recommended book of Boulos et al. [20], is useless for calculation of gas flow. Introduced in the early 1900s, the Hazen–Williams equation determines pipe friction head loss for water, requiring a single roughness coefficient (roughness is also very important parameter also in Darcy-Weisbach scheme for calculation [36]). Unfortunately even for water it may produce errors as high as ±40% when applied outside a limited and somewhat controversial range of the Reynolds numbers, pipe diameters and coefficients. Not only inaccurate the Hazen-Williams equation is conceptually incorrect [37].

In this paper the focus is on pipes, and other parts of systems are not examined. Furthermore, in a water or gas distribution system, the pipe friction head losses usually predominate and other minor losses can be ordinarily neglected without serious errors [38-41].

## 5. Topology of looped pipe system

First of all, maximal consumption per each node including one or more inlet nodes has to be determined (red in Figure 2). These parameters are looked up during the calculation. Further, initial guess of flow per conduits has to be assigned to satisfy first Kirchhoff's law and in that way chosen values are to be used for first iteration [3]. Final flows do not depend on first assumed flows per pipes (countless initial flow pattern can satisfy first Kirchhoff's law and all of them equally can be used with the same final results [3,38]). After the iteration procedure is completed, and if the value of gas or water flow velocity for all conduits are bellow standard values, calculated flows become flow distribution per pipes for maximal possible consumptions per nodes. Further, pressure per all nodes (can be heads in case of water) can be calculated. Whole network can be supplied by gas or water from one or more points (nodes). Distribution network must be design for largest consumption assigned to nodes of networks chosen to satisfy larges possible gas i.e. water consumption of households. Disposal of households is along the network's conduits, and only their consumption is to be assigned to nodes. Main purpose of the method is to calculate flow pattern per pipes and pressure pattern per nodes for the maximal load of the network[1]. First assumed flow pattern has to be chosen to satisfy first Kirchhoff's law (continuity of flow) which means that algebraic sum of flows per each node must be zero exactly. On the other hand, second Kirchhoff's law (continuity of potential), which means that algebraic sum of pressure drops per each contour must be approximately zero at the end of iterative procedure. Procedure can be interrupted when algebraic sum per all nodes become approximately zero, or when flows per pipes are not changed in calculation after two successive iterations.

One spatial fluid distribution network of pipelines will be examined as example (Figure 2). Polyethylene pipes (PVC) are used in the example shown in this paper.

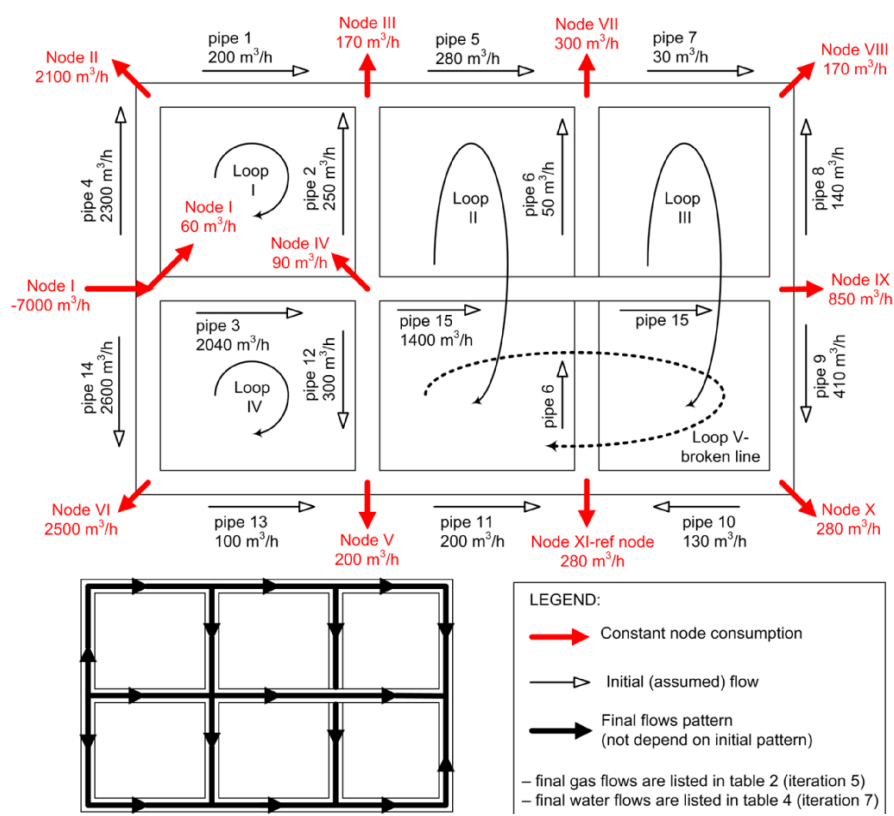

**Figure 2.** Spatial gas/water distribution network with loops – example

The first step in solving a problem is to make a network map showing pipe sizes and lengths, connections between pipes (nodes), and sources of supply. For convenience in locating pipes, assign each loop and each pipe a code number. Some of the pipes are mutual to one loop and some to two or even three contours (i.e. pipe 12 belongs to the loops II, IV, V). Special cases may occur in which two pipes cross each other but are not connected (like pipes 6 and 15), resulting in certain pipes being common to three or more loops. The distribution network then becomes three-dimensional (rare for gas with exception of maybe some chemical engineering facilities, water networks or district heating system, and usually for airflow networks). For example, loop V consists of conduits 15, 9, 10, via 11, and 12. Gas/water flow into the network from a source on the left side is 7000 m³/h, and points of delivery are at junctions of pipes (nodes), with the red arrows pointing to volumes delivered (node consumption). Summation of these deliveries equals 7000 m³/h. Assumed gas flows and their directions are indicated by black arrows near the pipes (Figure 2).

---

[1] Problem can be treated as inverse, i.e. flow per pipes assigned in the first iteration is not only initial pattern see (17). This flow pattern is not variable in further calculation. Instead of flows per pipe which are now constants, pipe diameters become variables, and according to this approach, optimized pipes' diameters in the network are the final result of calculation (see section 8 of this paper).

## 6. Topology equations for the observed looped network of pipes

After the network map with its pipes and loop numbers and delivery and supply data has been prepared, mathematical description of the network can be done. To introduce matrix form in calculation, it is necessary to represent distribution network from Figure 2 as a graph according to Euler's theorem from mineralogy (number of polyhedral angles and edges of minerals). Graph has X branches and Y nodes where in Figure 2, X = 15 and Y = 11). Graph with n nodes (in our case 11) has Y-1 independent nodes (in our case 10) and X-Y+1 independent loops (in our case 5). Tree is a set of connected branches chosen to connect all nodes, but not to make any closed path (not to form a loop). Branches, which do not belong to a tree, are links (number of links are X-Y+1). Loops in the network are formed using pipes from tree and one more chosen among the link pipes). Number of the loops is determined by number of links. In graph, one node is referent[2] (in Figure 2 referent node is XI) and all others are so called dependent nodes.

*6.1. Loop equations*

The Renouard equation (1) will be used for calculation of pressure drop in pipes in the case of natural gas distribution [31].

$$F_g = \Delta \tilde{p}^2 = p_1^2 - p_2^2 = 4810 \cdot \frac{\rho_r \cdot L \cdot Q^{1.82}}{\delta^{4.82}} \tag{1}$$

Regarding to Renouard formula (1) one has to be careful since it does not relate pressure drop but actually difference of the quadratic pressure at the input and the output of conduit. This means that $\sqrt{F}$ is not actually pressure drop in spite of the same unit of measurement, i.e. same unit is used as for pressure (Pa). Parameter $\sqrt{F}$ rather can be noted as pseudo-pressure drop. Fact that when $\sqrt{F} \to 0$ this consecutive means that also $F \to 0$ is very useful for calculation of gas pipeline with loops. So, notation for pseudo-pressure drop $\Delta p^2$ is ambiguous [3] (only $F$ or $\Delta \tilde{p}^2$ with appropriate index should be used instead of $\Delta p^2$).

First derivative of previous relation where the flow is treated as variable is (2):

$$F'_g = \frac{\partial F_g(Q)}{\partial Q} = 1.82 \cdot 4810 \cdot \frac{\rho_r \cdot L \cdot Q^{0.82}}{\delta^{4.82}} \tag{2}$$

---

[2] In other approach, with no referent node, one pseudo-loop must be introduced [39]. This is very complicated and should be avoided.

The Colebrook-White equation (3) will be used for calculation the Darcy's friction factor in the case of water distribution [32]. The Colebrook-White equation is implicit in friction factor, and here it is solved using MS Excel.

$$\frac{1}{\sqrt{\lambda}} = -2 \cdot \log_{10}\left(\frac{2.51}{Re} \cdot \frac{1}{\sqrt{\lambda}} + \frac{\varepsilon}{3.71 \cdot \delta}\right) \qquad (3)$$

Friction factor $\lambda$ calculated after Colebrook's relation will be incorporated into the Darcy-Weisbach relation to calculate pressure drop in water network (4).

$$F_w = \Delta p = p_1 - p_2 = \lambda \cdot \frac{L}{\delta^5} \cdot \frac{8 \cdot Q^2}{\pi^2} \cdot \rho \qquad (4)$$

Similar as for the gas lines, first derivate of previous relation where the flow is treated as variable is (5):

$$F'_w = \frac{\partial F_w(Q)}{\partial Q} = \lambda \cdot \frac{L}{\delta^5} \cdot \frac{16 \cdot Q}{\pi^2} \cdot \rho_{,,} \qquad (5)$$

Then, according to previous, for the gas network from figure 2, set of loop equation can be written as (6):

$$\left.\begin{array}{c}
\Delta \tilde{p}_1^2 - \Delta \tilde{p}_2^2 - \Delta \tilde{p}_3^2 + \Delta \tilde{p}_4^2 = \\
= 4810 \cdot \rho_r \left(\frac{L_1 \cdot Q_1^{1.82}}{\delta_1^{4.82}} - \frac{L_2 \cdot Q_2^{1.82}}{\delta_2^{4.82}} - \frac{L_3 \cdot Q_3^{1.82}}{\delta_3^{4.82}} + \frac{L_4 \cdot Q_4^{1.82}}{\delta_4^{4.82}}\right) \\
\Delta \tilde{p}_2^2 + \Delta \tilde{p}_5^2 - \Delta \tilde{p}_6^2 - \Delta \tilde{p}_{11}^2 + \Delta \tilde{p}_{12}^2 = \\
= 4810 \cdot \rho_r \left(\frac{L_2 \cdot Q_2^{1.82}}{\delta_2^{4.82}} + \frac{L_5 \cdot Q_5^{1.82}}{\delta_5^{4.82}} - \frac{L_6 \cdot Q_6^{1.82}}{\delta_6^{4.82}} - \frac{L_{11} \cdot Q_{11}^{1.82}}{\delta_{11}^{4.82}} + \frac{L_{12} \cdot Q_{12}^{1.82}}{\delta_{12}^{4.82}}\right) \\
\Delta \tilde{p}_6^2 + \Delta \tilde{p}_7^2 - \Delta \tilde{p}_8^2 + \Delta \tilde{p}_9^2 + \Delta \tilde{p}_{10}^2 = \\
= 4810 \cdot \rho_r \left(\frac{L_6 \cdot Q_6^{1.82}}{\delta_6^{4.82}} + \frac{L_7 \cdot Q_7^{1.82}}{\delta_7^{4.82}} - \frac{L_8 \cdot Q_8^{1.82}}{\delta_8^{4.82}} + \frac{L_{11} \cdot Q_9^{1.82}}{\delta_9^{4.82}} + \frac{L_{10} \cdot Q_{10}^{1.82}}{\delta_{10}^{4.82}}\right) \\
\Delta \tilde{p}_3^2 + \Delta \tilde{p}_{12}^2 - \Delta \tilde{p}_{13}^2 - \Delta \tilde{p}_{14}^2 = \\
= 4810 \cdot \rho_r \left(\frac{L_3 \cdot Q_3^{1.82}}{\delta_3^{4.82}} + \frac{L_{12} \cdot Q_{12}^{1.82}}{\delta_{12}^{4.82}} - \frac{L_{13} \cdot Q_{13}^{1.82}}{\delta_{13}^{4.82}} - \frac{L_{14} \cdot Q_{14}^{1.82}}{\delta_{14}^{4.82}}\right) \\
\Delta \tilde{p}_9^2 + \Delta \tilde{p}_{10}^2 - \Delta \tilde{p}_{11}^2 - \Delta \tilde{p}_{12}^2 + \Delta \tilde{p}_{15}^2 = \\
= 4810 \cdot \rho_r \left(\frac{L_9 \cdot Q_9^{1.82}}{\delta_9^{4.82}} + \frac{L_{10} \cdot Q_{10}^{1.82}}{\delta_{10}^{4.82}} - \frac{L_{11} \cdot Q_{11}^{1.82}}{\delta_{11}^{4.82}} - \frac{L_{12} \cdot Q_{12}^{1.82}}{\delta_{12}^{4.82}} + \frac{L_{15} \cdot Q_{15}^{1.82}}{\delta_{15}^{4.82}}\right)
\end{array}\right\} \begin{array}{l} F_I \\ F_{II} \\ F_{III} \\ F_{IV} \\ F_V \end{array} loops \qquad (6)$$

Previous relations can be noted in matrix form as (7):

$$\begin{bmatrix} 1 & -1 & -1 & 1 & 0 & 0 & 0 & 0 & 0 & 0 & 0 & 0 & 0 & 0 & 0 \\ 0 & 1 & 0 & 0 & 1 & -1 & 0 & 0 & 0 & 0 & -1 & -1 & 0 & 0 & 0 \\ 0 & 0 & 0 & 0 & 0 & 1 & 1 & -1 & 1 & 1 & 0 & 0 & 0 & 0 & 0 \\ 0 & 0 & 1 & 0 & 0 & 0 & 0 & 0 & 0 & 0 & 1 & -1 & -1 & 0 \\ 0 & 0 & 0 & 0 & 0 & 0 & 0 & 0 & 1 & 1 & -1 & -1 & 0 & 0 & 1 \end{bmatrix} x \begin{bmatrix} \Delta \tilde{p}_1^2 \\ \Delta \tilde{p}_2^2 \\ \Delta \tilde{p}_3^2 \\ \vdots \\ \Delta \tilde{p}_{15}^2 \end{bmatrix} = 0, \qquad (7)$$

Or for waterworks or district heating systems from figure 2 can be noted as (8):

$$\left.\begin{aligned}
&\Delta p_1 - \Delta p_2 - \Delta p_3 + \Delta p_4 = \\
&= \frac{8 \cdot \rho}{\pi^2} \cdot \left(\frac{\lambda_1 \cdot L_1 \cdot Q_1^2}{\delta_1^5} - \frac{\lambda_2 \cdot L_2 \cdot Q_2^2}{\delta_2^5} - \frac{\lambda_3 \cdot L_3 \cdot Q_3^2}{\delta_3^5} + \frac{\lambda_4 \cdot L_4 \cdot Q_4^2}{\delta_4^5}\right) \\
&\Delta p_2 + \Delta p_5 - \Delta p_6 - \Delta p_{11} + \Delta p_{12} = \\
&= \frac{8 \cdot \rho}{\pi^2} \cdot \left(\frac{\lambda_2 \cdot L_2 \cdot Q_2^2}{\delta_2^5} + \frac{\lambda_5 \cdot L_5 \cdot Q_5^2}{\delta_5^5} - \frac{\lambda_6 \cdot L_6 \cdot Q_6^2}{\delta_6^5} - \frac{\lambda_{11} \cdot L_{11} \cdot Q_{11}^2}{\delta_{11}^5} + \frac{\lambda_{12} \cdot L_{12} \cdot Q_{12}^2}{\delta_{12}^5}\right) \\
&\Delta p_6 + \Delta p_7 - \Delta p_8 + \Delta p_9 + \Delta p_{10} = \\
&= \frac{8 \cdot \rho}{\pi^2} \cdot \left(\frac{\lambda_6 \cdot L_6 \cdot Q_6^2}{\delta_6^5} + \frac{\lambda_7 \cdot L_7 \cdot Q_7^2}{\delta_7^5} - \frac{\lambda_8 \cdot L_8 \cdot Q_8^2}{\delta_8^5} + \frac{\lambda_9 \cdot L_9 \cdot Q_9^2}{\delta_9^5} + \frac{\lambda_{10} \cdot L_{10} \cdot Q_{10}^2}{\delta_{10}^5}\right) \\
&\Delta p_3 + \Delta p_{12} - \Delta p_{13} - \Delta p_{14} = \\
&= \frac{8 \cdot \rho}{\pi^2} \cdot \left(\frac{\lambda_3 \cdot L_3 \cdot Q_3^2}{\delta_3^5} + \frac{\lambda_{12} \cdot L_{12} \cdot Q_{12}^2}{\delta_{12}^5} - \frac{\lambda_{13} \cdot L_{13} \cdot Q_{13}^2}{\delta_{13}^5} - \frac{\lambda_{14} \cdot L_{14} \cdot Q_{14}^2}{\delta_{14}^5}\right) \\
&\Delta p_9 + \Delta p_{10} - \Delta p_{11} - \Delta p_{12} + \Delta p_{15} = \\
&= \frac{8 \cdot \rho}{\pi^2} \cdot \left(\frac{\lambda_9 \cdot L_9 \cdot Q_9^2}{\delta_9^5} + \frac{\lambda_{10} \cdot L_{10} \cdot Q_{10}^2}{\delta_{10}^5} - \frac{\lambda_{11} \cdot L_{11} \cdot Q_{11}^2}{\delta_{11}^5} - \frac{\lambda_{12} \cdot L_{12} \cdot Q_{12}^2}{\delta_{12}^5} + \frac{\lambda_{15} \cdot L_{15} \cdot Q_{15}^2}{\delta_{15}^5}\right)
\end{aligned}\right\} \begin{matrix} F_I \\ F_{II} \\ F_{III} \\ F_{IV} \\ F_V \end{matrix} \text{loops} \quad (8)$$

i.e. in matrix form for water distribution (9).

$$\begin{bmatrix} 1 & -1 & -1 & 1 & 0 & 0 & 0 & 0 & 0 & 0 & 0 & 0 & 0 & 0 & 0 \\ 0 & 1 & 0 & 0 & 1 & -1 & 0 & 0 & 0 & 0 & -1 & -1 & 0 & 0 & 0 \\ 0 & 0 & 0 & 0 & 0 & 1 & 1 & -1 & 1 & 1 & 0 & 0 & 0 & 0 & 0 \\ 0 & 0 & 1 & 0 & 0 & 0 & 0 & 0 & 0 & 0 & 0 & 1 & -1 & -1 & 0 \\ 0 & 0 & 0 & 0 & 0 & 0 & 0 & 0 & 1 & 1 & -1 & -1 & 0 & 0 & 1 \end{bmatrix} x \begin{bmatrix} \Delta p_1 \\ \Delta p_2 \\ \Delta p_3 \\ \vdots \\ \Delta p_{15} \end{bmatrix} = 0, \quad (9)$$

In the left matrix of the relations (7) and (9) rows represent loops and columns represent pipes. These relations are matrix reformulation of the second Kirchhoff's law. The sign for the term relates if the assumed flow is clockwise (1) or counter-clockwise (-1) relative to the loop.

*6.2. Node equations*

For all nodes in the network from figure 2, relations after the first Kirchhoff's law can be noted as (10):

$$\left.\begin{aligned}
&-Q_3 - Q_4 - Q_{14} - Q_{I-\text{output}} + Q_{I-\text{input}} = 0 & \text{node}_I \\
&-Q_1 + Q_4 - Q_{II-\text{output}} = 0 & \text{node}_{II} \\
&Q_1 + Q_2 - Q_5 - Q_{III-\text{output}} = 0 & \text{node}_{III} \\
&-Q_2 + Q_3 - Q_{12} - Q_{15} - Q_{IV-\text{output}} = 0 & \text{node}_{IV} \\
&-Q_{11} + Q_{12} + Q_{13} - Q_{V-\text{output}} = 0 & \text{node}_V \\
&-Q_{13} + Q_{14} - Q_{VI-\text{output}} = 0 & \text{node}_{VI} \\
&Q_5 + Q_6 - Q_7 - Q_{VII-\text{output}} = 0 & \text{node}_{VII} \\
&Q_7 + Q_8 - Q_{VIII-\text{output}} = 0 & \text{node}_{VIII} \\
&-Q_8 - Q_9 + Q_{15} - Q_{IX-\text{output}} = 0 & \text{node}_{IX} \\
&Q_9 - Q_{10} - Q_{X-\text{output}} = 0 & \text{node}_X \\
&-Q_6 + Q_{10} + Q_{11} - Q_{XI-\text{output}} = 0 & \text{node}_{XI} - \text{ref}
\end{aligned}\right. \quad (10)$$

Or in matrix form as (11) where in the first matrix rows represents nodes excluding referent node[3]. The node matrix with all node included are not linearly independent. To obtain linear independence any row of the node matrix has to be omitted. No information on the topology in that way will be lost [22].

---

[3] Formulation where node 1 is the referent node see in Brkić [3]

$$\begin{bmatrix} 0 & 0 & -1 & -1 & 0 & 0 & 0 & 0 & 0 & 0 & 0 & 0 & 0 & -1 & 0 \\ -1 & 0 & 0 & 1 & 0 & 0 & 0 & 0 & 0 & 0 & 0 & 0 & 0 & 0 & 0 \\ 1 & 1 & 0 & 0 & -1 & 0 & 0 & 0 & 0 & 0 & 0 & 0 & 0 & 0 & 0 \\ 0 & -1 & 1 & 0 & 0 & 0 & 0 & 0 & 0 & 0 & -1 & 0 & 0 & -1 \\ 0 & 0 & 0 & 0 & 0 & 0 & 0 & 0 & 0 & -1 & 1 & 1 & 0 & 0 \\ 0 & 0 & 0 & 0 & 0 & 0 & 0 & 0 & 0 & 0 & 0 & -1 & 1 & 0 \\ 0 & 0 & 0 & 0 & 1 & 1 & -1 & 0 & 0 & 0 & 0 & 0 & 0 & 0 \\ 0 & 0 & 0 & 0 & 0 & 0 & 0 & -1 & -1 & 0 & 0 & 0 & 0 & 1 \\ 0 & 0 & 0 & 0 & 0 & 0 & 0 & 1 & -1 & 0 & 0 & 0 & 0 & 0 \\ 0 & 0 & 0 & 0 & 0 & -1 & 0 & 0 & 0 & 1 & 1 & 0 & 0 & 0 \end{bmatrix} * \begin{bmatrix} Q_1 \\ Q_2 \\ Q_3 \\ Q_4 \\ Q_5 \\ Q_6 \\ Q_7 \\ Q_8 \\ Q_9 \\ Q_{10} \\ Q_{11} \\ Q_{12} \\ Q_{13} \\ Q_{14} \\ Q_{15} \end{bmatrix} = \begin{bmatrix} Q_{I-output} - |Q_{I-input}| \\ Q_{II-output} \\ Q_{III-output} \\ Q_{IV-output} \\ Q_{V-output} \\ Q_{VI-output} \\ Q_{VII-output} \\ Q_{VIII-output} \\ Q_{IX-output} \\ Q_{X-output} \end{bmatrix} \quad (11)$$

First row corresponds to the first node, etc. Last row is for node 10 from figure 2, since the node 11 is chosen to be referent node and therefore must be omitted from the matrix. For example, node 1 has connection with other nodes via pipes 3, 4 and 14, and for first assumed flow pattern, all flows are from node 1 via connected pipes to other nodes. Therefore, terms 3, 4, and 14 in first row are -1. Other pipes are not connected with node 1, and therefore all other terms in the first row of node matrix are 0.

Note that there is no difference in cases of water apropos gas calculation when the node equations are observed.

## 7. Network calculation according to the node-loop method

The nodes and the loops equations shown in previous text here will be united in one coherent system by coupling these two set of equations. This method will be examined in details for the network shown in Figure 2. This network will be treated as natural gas network in the sections 7.1 and as water network in 7.2. This approach also gives good insight into the differences which can be occurred in the cases of distribution of liquids apropos gaseous fluids.

*7.1. The node-loop calculation of gas networks*

First iteration for the gas calculation for the network from Figure 2 is shown in Table 1. If sign of calculated flow is negative, this means that flow direction from previous iteration must be changed, otherwise, sing must be remained unchanged. In Table 1, loop and the pipes numbers are listed in the first and the second column, respectively. Pipe length expressed in meters is listed in the third column, and assumed gas flow in each pipe expressed in m$^3$/s is listed in the fourth column. The 1 or -1 in fifth column indicates sing preceding flow in the fourth column. The plus or minus preceding the flow, Q, indicates the direction of the pipe flow for the particular loop. A plus sign denotes clockwise flow in the pipe within the loop, a minus sign counterclockwise. All these assumption will not be changed also in the case of waterworks or district heating system calculation.

Table 1. Node-loop analysis for the gas network from Figure 1.

| Loop | Pipe | δ (m) | L (m) | [a]Q(m³/s) | Sign (Q) | [c]F | [d]\|F'\| |
|---|---|---|---|---|---|---|---|
| I | 1 | 0.4064 | 100 | [b]$A_1$=0.0556 | +1 | 114959 | \|$a_1$\|=3766062 |
|  | 2 | 0.3048 | 100 | $A_2$=-0.0694 | -1 | -690438 | \|$a_2$\|=18094990 |
|  | 3 | 0.1524 | 100 | $A_3$=-0.5667 | -1 | -889949040 | \|$a_3$\|=2858306918 |
|  | 4 | 0.3048 | 100 | $A_4$=0.6389 | +1 | 39193885 | \|$a_4$\|=111651451 |
|  |  |  |  |  | Σ | A=-851330634 |  |
| II | 5 | 0.1524 | 100 | $B_1$=0.0778 | +1 | 23969880 | \|$b_1$\|=560895181 |
|  | 6 | 0.3048 | 200 | $B_2$=-0.0139 | -1 | -73795 | \|$b_2$\|=9670144 |
|  | 11 | 0.1524 | 100 | $B_3$=-0.0556 | -1 | -12993101 | \|$b_3$\|=425654001 |
|  | 12 | 0.1524 | 100 | $B_4$=-0.0833 | -1 | -27176838 | \|$b_4$\|=593542132 |
|  | 2 | 0.3048 | 100 | $B_5$=0.0694 | +1 | 690438 | \|$b_5$\|=18094990 |
|  |  |  |  |  | Σ | B=-15583417 |  |
| III | 7 | 0.1524 | 100 | $C_1$=0.0083 | +1 | 411338 | \|$c_1$\|=89836237 |
|  | 8 | 0.1524 | 100 | $C_2$=-0.0389 | -1 | -6788773 | \|$c_2$\|=317714556 |
|  | 9 | 0.3048 | 100 | $C_3$=0.1139 | +1 | 1698792 | \|$c_3$\|=27147529 |
|  | 10 | 0.1524 | 100 | $C_4$=0.0361 | +1 | 5932191 | \|$c_4$\|=298982433 |
|  | 6 | 0.3048 | 200 | $C_5$=0.0139 | +1 | 73795 | \|$c_5$\|=9670144 |
|  |  |  |  |  | Σ | C=1327344 |  |
| IV | 3 | 0.1524 | 100 | $D_1$=0.5667 | +1 | 889949040 | \|$d_1$\|=2858306918 |
|  | 12 | 0.1524 | 100 | $D_2$=0.0833 | +1 | 27176838 | \|$d_2$\|=593542132 |
|  | 13 | 0.1524 | 100 | $D_3$=-0.0278 | -1 | -3679919 | \|$d_3$\|=241108279 |
|  | 14 | 0.4064 | 100 | $D_4$=-0.7222 | -1 | -12243919 | \|$d_4$\|=30854675 |
|  |  |  |  |  | Σ | D=901202040 |  |
| V | 15 | 0.1524 | 200 | $E_1$=0.3889 | +1 | 897059511 | \|$e_1$\|=4198238510 |
|  | 9 | 0.3048 | 100 | $E_2$=0.1139 | +1 | 1698792 | \|$e_2$\|=27147529 |
|  | 10 | 0.1524 | 100 | $E_3$=0.0361 | +1 | 5932191 | \|$e_3$\|=298982433 |
|  | 11 | 0.1524 | 100 | $E_4$=-0.0556 | -1 | -12993101 | \|$e_4$\|=425654001 |
|  | 12 | 0.1524 | 100 | $E_5$=-0.0833 | -1 | -27176838 | \|$e_5$\|=593542132 |
|  |  |  |  |  | Σ | E=864520555 |  |

[a]from Figure 2 but expressed in m³/s

[b]letters used in (13) and (14)

[c]see (1)

[d]see (2)

To introduce matrix calculation, the node-loop matrix [NL], matrix of calculated flow in observed iteration [Q], and [V] matrix in the right side of (12) will be defined.

$$[NL] \times [Q] = [V], \quad (12)$$

First ten rows in the NL (13) matrix are from node matrix (11), and next five rows are loop matrix (7 and 9). These five rows from the loop matrix are multiplied by first derivate of pressure drop function (2) from Table 1 for gas[4] (column F').

$$[NL] = \begin{bmatrix} 0 & 0 & -1 & -1 & 0 & 0 & 0 & 0 & 0 & 0 & 0 & 0 & 0 & -1 & 0 \\ -1 & 0 & 0 & 1 & 0 & 0 & 0 & 0 & 0 & 0 & 0 & 0 & 0 & 0 & 0 \\ 1 & 1 & 0 & 0 & -1 & 0 & 0 & 0 & 0 & 0 & 0 & 0 & 0 & 0 & 0 \\ 0 & -1 & 1 & 0 & 0 & 0 & 0 & 0 & 0 & 0 & 0 & -1 & 0 & 0 & -1 \\ 0 & 0 & 0 & 0 & 0 & 0 & 0 & 0 & 0 & 0 & -1 & 1 & 1 & 0 & 0 \\ 0 & 0 & 0 & 0 & 0 & 0 & 0 & 0 & 0 & 0 & 0 & -1 & 1 & 0 \\ 0 & 0 & 0 & 0 & 1 & 1 & -1 & 0 & 0 & 0 & 0 & 0 & 0 & 0 & 0 \\ 0 & 0 & 0 & 0 & 0 & 0 & 0 & -1 & -1 & 0 & 0 & 0 & 0 & 0 & 1 \\ 0 & 0 & 0 & 0 & 0 & 0 & 0 & 1 & -1 & 0 & 0 & 0 & 0 & 0 & 0 \\ 0 & 0 & 0 & 0 & 0 & -1 & 0 & 0 & 1 & 1 & 0 & 0 & 0 & 0 & 0 \\ 1 \cdot |a_1| & -1 \cdot |a_2| & -1 \cdot |a_3| & 1 \cdot |a_4| & 0 & 0 & 0 & 0 & 0 & 0 & 0 & 0 & 0 & 0 & 0 \\ 0 & 1 \cdot |b_5| & 0 & 0 & 1 \cdot |b_1| & -1 \cdot |b_2| & 0 & 0 & 0 & 0 & -1 \cdot |b_3| & -1 \cdot |b_4| & 0 & 0 & 0 \\ 0 & 0 & 0 & 0 & 0 & 1 \cdot |c_5| & 1 \cdot |c_1| & -1 \cdot |c_2| & 1 \cdot |c_3| & 1 \cdot |c_4| & 0 & 0 & 0 & 0 & 0 \\ 0 & 0 & 1 \cdot |d_1| & 0 & 0 & 0 & 0 & 0 & 0 & 0 & 0 & 1 \cdot |d_2| & -1 \cdot |d_3| & -1 \cdot |d_4| & 0 \\ 0 & 0 & 0 & 0 & 0 & 0 & 0 & 1 \cdot |e_2| & 1 \cdot |e_3| & -1 \cdot |e_4| & -1 \cdot |e_5| & 0 & 0 & 1 \cdot |e_1| \end{bmatrix} \quad (13)$$

First ten rows in matrix [V] are node consumption[5], and the rest five terms are from Table 1 (14).

$$[V] = \begin{bmatrix} Q_{I-output} - |Q_{I-input}| \\ Q_{II-output} \\ Q_{III-output} \\ Q_{IV-output} \\ Q_{V-output} \\ Q_{VI-output} \\ Q_{VII-output} \\ Q_{VIII-output} \\ Q_{IX-output} \\ Q_{X-output} \\ -A + (A_1 \cdot |a_1| + A_2 \cdot |a_2| + A_3 \cdot |a_3| + A_4 \cdot |a_4|) \\ -B + (B_1 \cdot |b_1| + B_2 \cdot |b_2| + B_3 \cdot |b_3| + B_4 \cdot |b_4| + B_5 \cdot |b_5|) \\ -C + (C_1 \cdot |c_1| + C_2 \cdot |c_2| + C_3 \cdot |c_3| + C_4 \cdot |c_4| + C_5 \cdot |c_5|) \\ -D + (D_1 \cdot |d_1| + D_2 \cdot |d_2| + D_3 \cdot |d_3| + D_4 \cdot |d_4|) \\ -E + (E_1 \cdot |e_1| + E_2 \cdot |e_2| + E_3 \cdot |e_3| + E_4 \cdot |e_4| + E_5 \cdot |e_5|) \end{bmatrix} \quad (14)$$

Solution of matrix [Q] is now (15):

$$[Q] = inv[NL] \times [V], \quad (15)$$

Sign minus in front of some term means that sing preceding this term from the previous iteration must be changed.

Five iterations are enough for the calculation of gas network from Figure 2. Calculated flows for these first five iterations will be listed in Table 2.

---

[4] For water (5) and Table 3

[5] Right side of (11), node consumptions (and input for node 1 with negative sign) from Figure 2 expressed in m³/s

Table 2. First five iteration for gas network from Figure 1 – example

| | Flow in m³/h | | | | | | cGas velocity |
|---|---|---|---|---|---|---|---|
| | Iteration | 1 | 2 | 3 | 4 | b5 | m/s |
| Pipe 1 | 200 | 687.38 | 1172.23 | 1225.74 | 1228.19 | 1228.19 | 0.66 |
| Pipe 2 | 250 | 33.55 | -307.01 | 360.38 | 362.80 | 362.80 | 0.35 |
| Pipe 3 | 2040 | 988.81 | 618.87 | 550.48 | 547.68 | 547.68 | 2.08 |
| Pipe 4 | 2300 | 2787.38 | 3272.23 | 3325.74 | 3328.19 | 3328.19 | 3.17 |
| Pipe 5 | 280 | 550.93 | 695.22 | 695.36 | 695.39 | 695.39 | 2.65 |
| Pipe 6 | 50 | 78.54 | -60.99 | 50.63 | 50.73 | 50.73 | 0.05 |
| Pipe 7 | 30 | 329.48 | 334.23 | 344.74 | 344.66 | 344.66 | 1.31 |
| Pipe 8 | 140 | -159.48 | 164.23 | 174.74 | 174.66 | 174.66 | 0.66 |
| Pipe 9 | 410 | 20.26 | -121.61 | 115.19 | 115.28 | 115.28 | 0.11 |
| Pipe 10 | 130 | -259.74 | 401.61 | 395.19 | 395.28 | 395.28 | 1.50 |
| Pipe 11 | 200 | 618.28 | 620.62 | 624.57 | 624.55 | 624.55 | 2.38 |
| Pipe 12 | 300 | 154.48 | 271.72 | 260.79 | 260.43 | 260.43 | 0.99 |
| Pipe 13 | 100 | 663.80 | 548.90 | 563.78 | 564.13 | 564.13 | 2.15 |
| Pipe 14 | 2600 | 3163.80 | 3048.90 | 3063.78 | 3064.13 | 3064.13 | 1.64 |
| Pipe 15 | 1400 | 710.78 | 564.16 | 560.07 | 560.05 | 560.05 | 2.13 |

aFirst assumed flows per pipes chosen after the first Kirchhoff's law (black letters in figure 2)

bValues in iterations 5 are equal as in iteration 4, stopping criterion is fulfilled

cGas velocity (10-15 m/s recommended); $\upsilon=(4 \cdot p \cdot Q)/(\delta^2 \cdot \pi)$; where p= $p_n/p_a$=0.25 and $p_a$ is absolute pressure of gas in the pipeline, here $p_a$=400 000 Pa, and $p_n$=normal pressure ~100 000 Pa, p=400 000 Pa/100 000 Pa=1/4

Flow direction is changed in pipe 2, 6, 8, 9 and 10 (opposite than assumed in first assumed flows). Note that the velocities in the last column of Table 2 are listed. Gas pressure in the network is circa 4x10⁵ Pa abs. Flow velocity per pipes is not balanced, somewhere is too small somewhere is too high. Whole problem can be treaded now as inverse by fixing the flows per pipes and optimized pipe diameters as noted in section 4. This can be done using here presented node-loop method, Hardy Cross or similar available methods (note that different pressure in the gas apropos water network causes different values of speed of gas compared to speed of water; last column in Table 2 and 4, respectively).

*7.2. The node-loop calculation of waterworks or district heating systems*

Similar as for gas networks, network from Figure 2 will be used for water distribution calculation (Table 3). The calculated flows listed in Table 4 are slightly different than for the gas flow calculation.

**Table 3.** Node-loop analysis for the water network from Figure 2.

| Loop | Pipe | δ (m) | L (m) | [a]Q(m³/s) | Sign (Q) | [c]Re | [d]ε/δ | [e]λ | [f]F | [g]|F'| |
|---|---|---|---|---|---|---|---|---|---|---|
| I | 1 | 0.4064 | 100 | [b]$A_1$=0.0556 | +1 | 195566.25 | 4.92·10⁻⁵ | 0.01609 | 363.1919278 | |$a_1$|=13074.9094 |
| | 2 | 0.3048 | 100 | $A_2$=-0.0694 | -1 | 325943.75 | 6.56·10⁻⁵ | 0.01492 | -2217.677686 | |$a_2$|=63869.11737 |
| | 3 | 0.1524 | 100 | $A_3$=-0.5667 | -1 | 5319401.99 | 1.31·10⁻⁴ | 0.01290 | -4084603.502 | |$a_3$|=14416247.66 |
| | 4 | 0.3048 | 100 | $A_4$=0.6389 | +1 | 2998682.50 | 6.56·10⁻⁵ | 0.01184 | 148932.0282 | |$a_4$|=466222.0014 |
| | | | | | | | Σ | | A=-3937526 | |
| II | 5 | 0.1524 | 100 | $B_1$=0.0778 | +1 | 730114.00 | 1.31·10⁻⁴ | 0.01423 | 84860.18126 | |$b_1$|=2182118.947 |
| | 6 | 0.3048 | 200 | $B_2$=-0.0139 | -1 | 65188.75 | 6.56·10⁻⁵ | 0.01998 | -237.4945042 | |$b_2$|=34199.2086 |
| | 11 | 0.1524 | 100 | $B_3$=-0.0556 | -1 | 521510.00 | 1.31·10⁻⁴ | 0.01470 | -44732.90001 | |$b_3$|=1610384.4 |
| | 12 | 0.1524 | 100 | $B_4$=-0.0833 | -1 | 782265.00 | 1.31·10⁻⁴ | 0.01414 | -96832.35986 | |$b_4$|=2323976.637 |
| | 2 | 0.3048 | 100 | $B_5$=0.0694 | +1 | 325943.75 | 6.56·10⁻⁵ | 0.01492 | 2217.677686 | |$b_5$|=63869.11737 |
| | | | | | | | Σ | | B=-54725 | |
| III | 7 | 0.1524 | 100 | $C_1$=0.0083 | +1 | 78226.50 | 1.31·10⁻⁴ | 0.01954 | 1338.024663 | |$c_1$|=321125.9191 |
| | 8 | 0.1524 | 100 | $C_2$=-0.0389 | -1 | 365057.00 | 1.31·10⁻⁴ | 0.01531 | -22830.90776 | |$c_2$|=1174160.971 |
| | 9 | 0.3048 | 100 | $C_3$=0.1139 | +1 | 534547.75 | 6.56·10⁻⁵ | 0.01391 | 5557.748158 | |$c_3$|=97599.47985 |
| | 10 | 0.1524 | 100 | $C_4$=0.0361 | +1 | 338981.50 | 1.31·10⁻⁴ | 0.01545 | 19868.97118 | |$c_4$|=1100435.327 |
| | 6 | 0.3048 | 200 | $C_5$=0.0139 | +1 | 65188.75 | 6.56·10⁻⁵ | 0.01998 | 237.4945042 | |$c_5$|=34199.2086 |
| | | | | | | | Σ | | C=4171 | |
| IV | 3 | 0.1524 | 100 | $D_1$=0.5667 | +1 | 5319401.99 | 1.31·10⁻⁴ | 0.01290 | 4084603.502 | |$d_1$|=14416247.66 |
| | 12 | 0.1524 | 100 | $D_2$=0.0833 | +1 | 782265.00 | 1.31·10⁻⁴ | 0.01414 | 96832.35986 | |$d_2$|=2323976.637 |
| | 13 | 0.1524 | 100 | $D_3$=-0.0278 | -1 | 260755.00 | 1.31·10⁻⁴ | 0.01600 | -12174.73104 | |$d_3$|=876580.635 |
| | 14 | 0.4064 | 100 | $D_4$=-0.7222 | -1 | 2542361.25 | 4.92·10⁻⁵ | 0.01157 | -44129.48853 | |$d_4$|=122204.7375 |
| | | | | | | | Σ | | D=4125132 | |
| V | 15 | 0.1524 | 200 | $E_1$=0.3889 | +1 | 3650569.99 | 1.31·10$^{-4}$ | 0.01302 | 3882751.322 | |$e_1$|=19968435.37 |
| | 9 | 0.3048 | 100 | $E_2$=0.1139 | +1 | 534547.75 | 6.56·10$^{-5}$ | 0.01391 | 5557.748158 | |$e_2$|=97599.47985 |
| | 10 | 0.1524 | 100 | $E_3$=0.0361 | +1 | 338981.50 | 1.31·10$^{-4}$ | 0.01545 | 19868.97118 | |$e_3$|=1100435.327 |
| | 11 | 0.1524 | 100 | $E_4$=-0.0556 | -1 | 521510.00 | 1.31·10$^{-4}$ | 0.01470 | -44732.90001 | |$e_4$|=1610384.4 |
| | 12 | 0.1524 | 100 | $E_5$=-0.0833 | -1 | 782265.00 | 1.31·10$^{-4}$ | 0.01414 | -96832.35986 | |$e_5$|=2323976.637 |
| | | | | | | | Σ | | E=3766613 | |

[a]from Figure 2 but expressed in m$^3$/s, [b]letters used in (13) and (14), [c]Reynolds number; dynamic water viscosity 0.00089 Pa·s, [d]Relative roughness; absolute roughness ε=0.00002 m for PVC pipes, [e]Friction factor (3) calculated using MS Excel, [f]Pressure drop in pipe (4), [g]see (5)

**Table 4.** First seven iteration for water network from Figure 2 – example, Flow in m$^3$/h

| Iteration | | 1 | 2 | 3 | 4 | 5 | 6 | [b]7 | Water velocity, m/s |
|---|---|---|---|---|---|---|---|---|---|
| Pipe 1 | 200 | 619.22 | 1117.82 | 1205.89 | 1214.92 | 1215.25 | 1215.26 | 1215.26 | 2.6 |
| Pipe 2 | 250 | 69.21 | -260.68 | 345.80 | 354.68 | 355.00 | 355.01 | 355.01 | 1.4 |
| Pipe 3 | 2040 | 1071.47 | 671.88 | 567.12 | 556.60 | 556.22 | 556.21 | 556.21 | 8.5 |
| Pipe 4 | 2300 | 2719.22 | 3217.82 | 3305.89 | 3314.92 | 3315.25 | 3315.26 | 3315.26 | 12.6 |
| Pipe 5 | 280 | 518.43 | 687.14 | 690.09 | 690.24 | 690.25 | 690.25 | 690.25 | 10.5 |
| Pipe 6 | 50 | 90.95 | -57.70 | 43.41 | 43.11 | 43.10 | 43.10 | 43.10 | 0.2 |
| Pipe 7 | 30 | 309.38 | 329.44 | 346.68 | 347.13 | 347.15 | 347.15 | 347.15 | 5.3 |
| Pipe 8 | 140 | -139.38 | 159.44 | 176.68 | 177.13 | 177.15 | 177.15 | 177.15 | 2.7 |
| Pipe 9 | 410 | 47.60 | -115.49 | 113.24 | 113.39 | 113.39 | 113.39 | 113.39 | 0.4 |
| Pipe 10 | 130 | -232.40 | 395.49 | 393.24 | 393.39 | 393.39 | 393.39 | 393.39 | 6.0 |
| Pipe 11 | 200 | 603.35 | 617.79 | 629.83 | 630.28 | 630.29 | 630.29 | 630.29 | 9.6 |
| Pipe 12 | 300 | 154.04 | 267.49 | 262.84 | 261.80 | 261.76 | 261.76 | 261.76 | 4.0 |
| Pipe 13 | 100 | 649.31 | 550.30 | 566.99 | 568.48 | 568.53 | 568.54 | 568.54 | 8.7 |
| Pipe 14 | 2600 | 3149.31 | 3050.30 | 3066.99 | 3068.48 | 3068.53 | 3068.54 | 3068.54 | 6.6 |
| Pipe 15 | 1400 | 758.22 | 575.07 | 560.08 | 559.48 | 559.46 | 559.46 | 559.46 | 8.5 |

[a]First assumed flows per pipes chosen after the first Kirchhoff's law (black letters in figure 2), [b]Values in iterations 7 are equal as in iteration 6, stopping criterion is fulfilled

## 8. A note on optimization problem

Renouard formula (1) for condition in gas distribution networks assumes a constant density of a fluid within the conduits. This assumption applies only to incompressible, i.e. for liquids flows such as in water distribution systems for municipalities (or any other liquid, like crude oil, etc.). For the small pressure drops in typical gas distribution networks, gas density can be treated as constant, which means that gas can be treated as incompressible fluid. Assumption of gas incompressibility means that it is compressed and forced to convey through conduits, but inside the pipeline system pressure drop of already compressed gas is minor and hence further changes in gas density can be neglected. Fact is that gas is actually compressed and hence that volume of gas is decreased and then such compressed volume of gas is conveying with constant density through gas distribution pipeline. So, mass of gas is constant, but volume is decreased while gas density is according to this, increased. Operate pressure for typical distribution gas network is 4x10$^5$ Pa abs i.e. 3x10$^5$ Pa gauge and accordingly volume of gas is decreased four times compared to volume of gas at normal (standard) conditions. But operate pressure for gas distribution network can be lower (this case is valid for network in paper of Brkić [3]). This is not typical for natural gas distributive networks. This was common practice in obsolete systems for distribution of city gas derived from coal [30]. So, flow in Renourad formula (1) adjusted for natural gas is usually expressed in normal (standard) conditions. Consequence is that if flows in previous paper of Brkić [3] are expressed in their real (compressed) values and if these real values are numerically equalized with values expressed for normal (standard) conditions, this means that operate pressure in gas network is normal (standard). Otherwise, velocities in previous paper of Brkić [3] have to be corrected. Velocities in previous paper of Brkić [3] are calculated to be comparable with the procedure shown in Manojlović et al [40] where calculation of gas distribution network in Serbian town Kragujevac is discussed. In Manojlović et al [40], flows are expressed in their real values and not for normal or standard conditions of pressure as common practice is or this network is calculated to work with lower pressure typical for gasses derived from coal. Second assumption can be rejected as less possible because in the part of Serbia south of rivers Sava and Danube where Kragujevac is situated, such gas was never used and especially not in 1990's. Some comments about that issue was also shown in Brkić [5]. So, to avoid any further ambiguity, conclusion is that all flows previous paper of Brkić [3] are expressed in their real (compressed) values while operate pressure at the inputs of shown networks is normal (standard).

If these values of flows are noted for normal (standard) conditions of pressure as common practice is (Table 2), while operate pressure is 4x10$^5$ Pa abs i.e. 3x10$^5$ Pa gauge, velocities of gas are different than those in previous paper of Brkić [3] while flows remain unchanged.

Velocities in Table 2 are calculated using (16):

$$v = \frac{4 \cdot p_n \cdot Q_n}{p_a \cdot \delta^2 \cdot \pi} = \frac{4 \cdot Q}{\delta^2 \cdot \pi} \tag{16}$$

Now, for such values of flows, diameters of conduits are too large and in such case Hardy Cross method [1] as well as improved Hardy Cross method [2,3] can be used for optimization of diameters of conduits shown in Figure 2. In a problem of optimization of pipe diameters, in Renouard formula (1), flow is not any more treated as variable (17) while correction Δ is now correction of diameters.

$$\frac{\partial(p_1^2 - p_2^2)}{\partial \delta} = \frac{\partial F_g(\delta)}{\partial \delta} = \frac{\partial \left(\frac{4810 \cdot Q^{1.82} \cdot L \cdot \rho_r}{\delta^{4.82}}\right)}{\partial \delta} = \frac{-4.82 \cdot 4810 \cdot Q^{1.82} \cdot L \cdot \rho_r}{\delta^{5.82}} \tag{17}$$

Ambiguity related to pressure conditions in a gas distributive network can cause very different and large consequences in an interpretation of calculated results.

Similar analogy regarding to water networks is clear (18):

$$\frac{\partial(\Delta p)}{\partial \delta} = \frac{\partial F_w(\delta)}{\partial \delta} = \frac{\partial \left(\frac{8 \cdot \rho \cdot \lambda \cdot L \cdot Q^2}{\pi^2 \cdot \delta^5}\right)}{\partial \delta} = \frac{-5 \cdot 8 \cdot \rho \cdot \lambda \cdot L \cdot Q^2}{\pi^2 \cdot \delta^6} \tag{18}$$

Diameters of conduits in presented gas pipeline should be optimized while diameters in water network are in an accepted tolerance.

## 9. Conclusions

Here presented the node-loop method is powerful numerical procedure for calculation of flows or diameters as inverse problems in looped fluid distribution networks. Main advantages is that flow in each pipe can be calculated directly, which is not possible after Hardy Cross and improved Hardy Cross methods (Figure 3). Similar numbers of iterations are necessary to achieve demanded accuracy in calculation as in the modified Hardy Cross method[6] (Figure 4).

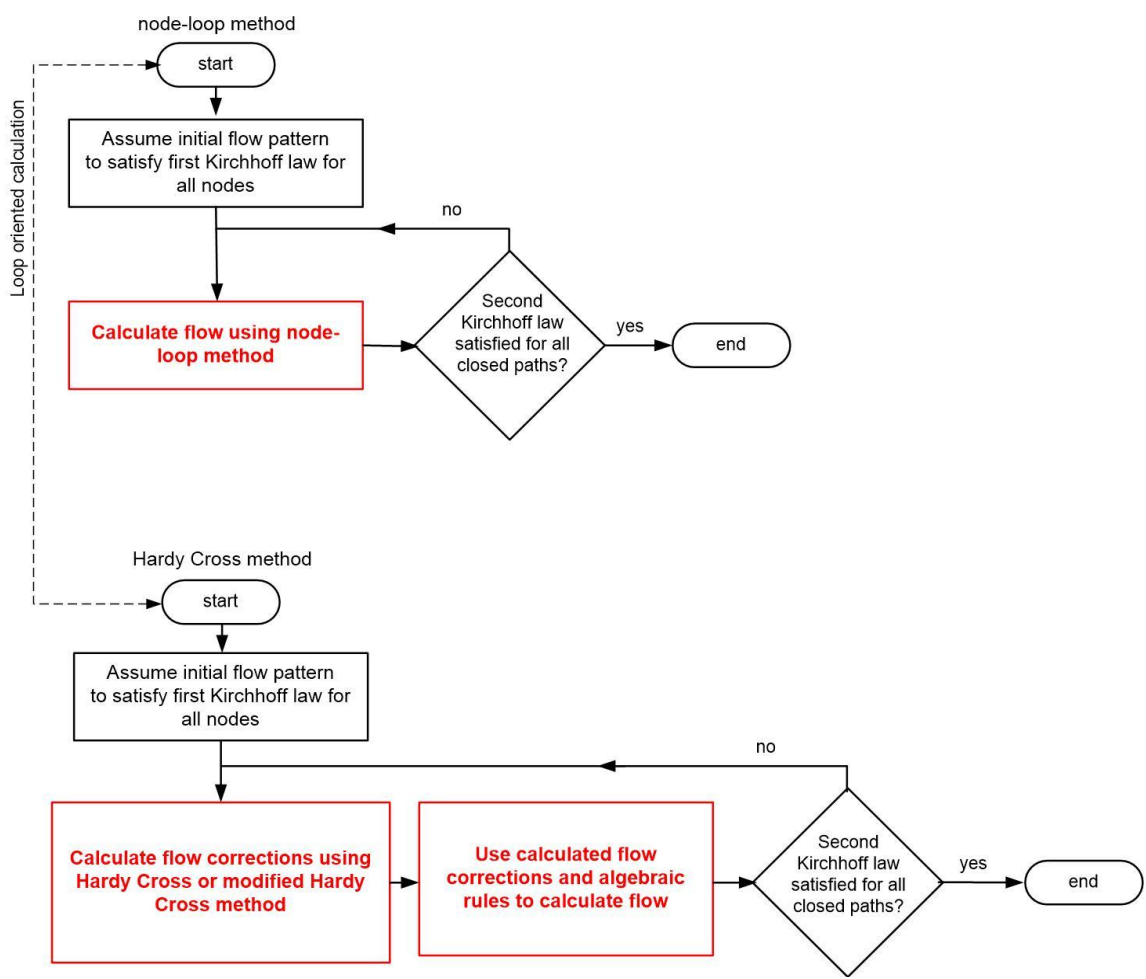

**Figure 3.** Main conceptual difference between the Hardy Cross method (original and improved) and the node-loop method

The hydraulic computations involved in designing water or gas distribution systems can be only approximated, as it is impossible to consider all the factors affecting loss of head in a complicated network of pipes.

The here presented methods can be easily readapted for detection of a position of leakage in a pipe network [42,43].

---

[6] Results for the Hardy Cross calculations are from the paper of Brkić [3]

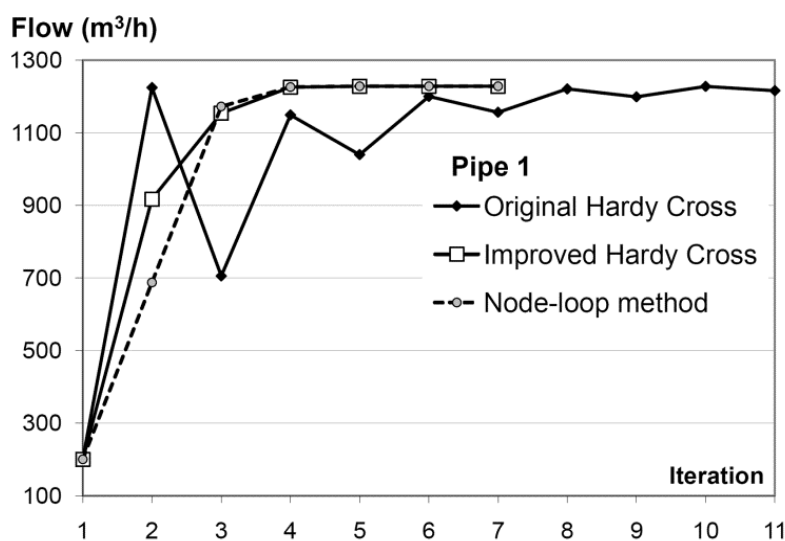

**Figure 4.** Comparison of the convergence performances for the Hardy Cross methods, original and improved, and the node-loop method

**Nomenclature**

p – pressure (Pa)
ϱr – relative gas density (-)
L – pipe length (m)
Q – fluid flow rate (m$^3$/s)
δ – pipe diameter (m)
Re – Reynolds number (-)
ε – absolute roughness of inner pipe surface (m)
ϱ – water density (kg/m$^3$)
υ – velocity (m/s)
λ – Darcy (i.e. Moody or Darcy-Weisbach) friction factor (-)
F – pressure function (Pa for water, and Pa$^2$ for natural gas)
$\Delta \tilde{p}$ - pseudo-pressure drop (Pa)

A to E and a to e – auxiliary symbols

Subscripts:
n – normal
w – water
g - gas
a – absolute

Constants:
π≈3.1415